\documentclass[conference]{IEEEtran}
\IEEEoverridecommandlockouts
\usepackage{cite}
\usepackage{amsmath,amssymb,amsfonts}
\usepackage{algorithmic}
\usepackage{graphicx}
\usepackage{textcomp}
\usepackage{xcolor}
\usepackage{booktabs}
\usepackage{xspace}
\usepackage{comment}
\usepackage[most]{tcolorbox}
\usepackage{tabularx}
\usepackage[table]{xcolor}
\usepackage{array}

\definecolor{headerblue}{RGB}{0, 0, 128}
\definecolor{summarygreen}{RGB}{0, 128, 0}
\definecolor{highred}{RGB}{180, 0, 0}
\definecolor{mediumgold}{RGB}{150, 150, 0}
\definecolor{lightgrey}{RGB}{245, 245, 245}

\newcolumntype{L}{>{\raggedright\arraybackslash}X}
\def\BibTeX{{\rm B\kern-.05em{\sc i\kern-.025em b}\kern-.08em
    T\kern-.1667em\lower.7ex\hbox{E}\kern-.125emX}}

\usepackage{hyperref}
\newcommand{\tool}{\textbf{D}eep\textbf{F}ix\xspace}
\newcommand{\as}[1]{\textcolor{green}{\textsf{\textbf{AS: #1}}}}
\begin{document}

\title{\tool: Debugging and Fixing Machine Learning Workflow using Agentic AI}

\author{\IEEEauthorblockN{1\textsuperscript{st} Fadel Mamar Seydou}
\IEEEauthorblockA{\textit{Technical University of Munich}\\
\textit{Delcaux Labs} \\
Stuttgart, Germany \\
fadel.seydou@tum.de}
\and
\IEEEauthorblockN{2\textsuperscript{nd} Arnab Sharma}
\IEEEauthorblockA{\textit{Heinz Nixdorf Institute} \\
\textit{Paderborn University}\\
Paderborn, Germany \\
arnab.sharma@uni-paderborn.de}
}

\maketitle

\begin{abstract}
In recent years, machine learning (ML) based software systems are increasingly deployed in several critical applications, yet systematic testing of their behavior remains challenging due to complex model architectures, large input spaces, and evolving deployment environments. Existing testing approaches often rely on generating test cases based on given requirements, which often fail to reveal critical bugs of modern ML models due to their complex nature. Most importantly, such approaches, although they can be used to detect the presence of specific failures in the ML software, they hardly provide any message as to how to fix such errors. To tackle this, in this paper, we present \tool, a tool for automated testing of the entire ML pipeline using an \textit{agentic AI} framework. Our testing approach first leverages Deepchecks to test the ML software for any potential bugs, and thereafter, uses an agentic AI-based approach to generate a detailed bug report. This includes a ranking, based on the severity of the found bugs, along with their explanations, which can be interpreted easily by any non-data science experts and most importantly, also provides possible ways to fix these bugs.
Additionally, \tool supports several \textit{types} of ML software systems and can be integrated easily to any ML workflow, enabling continuous testing throughout the development lifecycle. We discuss our already validated cases as well as some planned validations designed to demonstrate how the agentic testing process can reveal hidden failure modes that remain undetected by conventional testing methods. A 5-minute screencast demonstrating the tool's core functionality is available at \url{https://youtu.be/WfwZmFcQgBQ}.
\end{abstract}


\section{Introduction}

In traditional software systems, the failure can be defined as the explicit execution of crashes or the violation of a specific property. In contrast to typical software, however, the functionality of the 
ML-based software systems is {\em trained}. 
Specifically, the development of ML-based software systems involves complex data pipelines, large model architectures, and non-deterministic training algorithms~\cite{nguyen_systematic_2025}.
Therefore, the existing testing approaches for the traditional software systems cannot simply be applied to ML testing. To this end, the existing literature considered two different types of approaches (a) designing an ML algorithm to generate an ML model that is guaranteed to satisfy a specific requirement in ML software~\cite{ZhangCZX023}, (b) verifying or testing the requirements of the ML software~\cite{MurphySK09,sharma_mlcheck_2021,HumbatovaKJYT25}. 


Over the years, researchers have mainly focused on testing as a measure to ensure the quality of components of the ML workflow for further ensuring correctness of the ML software system. For instance, several works targeted the \emph{data} component and proposed methods to detect bugs in training data, labels, and preprocessing pipelines~\cite{GrafbergerGSS22}. Considering the implementation of the ML algorithm, \textit{metamorphic testing} approach has been shown to be quite effective in testing the implementation of the ML software~\cite{MurphySK09,SharmaW19,cho_metamorphic_2025}.
Apart from testing, formal verification methods have mainly focused on neural-network-specific properties, especially local robustness~\cite{CordeiroDGIJKKLMSW25}. Note that although such works can give the proof of specific properties, they are bound to a specific model and also computationally quite expensive.
There exist some works that focus on testing any type of ML models, i.e., model agnostic, such as~\cite{MurphySK09,sharma_mlcheck_2021,SharmaW20issta}. However, such approaches are limited by what they can test, i.e., such approaches allow only a specific type of properties that can be tested.  Most importantly,
Ngyuen et al.~\cite{nguyen_systematic_2025} argued that the research in ML testing remains fragmented across components such as data, training, architecture, and APIs, and that many practically relevant faults---especially in data processing and API usage---are still insufficiently addressed in the literature. This current ecosystem, considering fragmented testing approaches for the ML components forces practitioners to manually piece together clues from disjointed frameworks, significantly increasing their cognitive load \cite{schoop_umlaut_2021}. As a result, many developers lack comprehensive mental models of the training process and rely on trial-and-error debugging \cite{balayn_faulty_2023}\cite{schoop_umlaut_2021}. To this end, DeepCheck~\cite{deepchecks_open_source} provides a way to test the ML workflow, including the training data and the ML algorithm, with respect to specific functional properties. However, the bug report generated by Deepchecks cannot be easily interpreted unless one is familiar with the data science jargon. Most importantly, this bug report gives the information regarding the existence of the bugs, and does not provide any possible ways to fix them. 
Considering these factors, we asked ourselves the following question.

\begin{center}
\fbox{\parbox{0.8\linewidth}{\centering\textit{Can we develop an approach that could be used for
all the components of the ML workflow, and the bug report contains the possible fixes?}}}
\end{center}

To provide a practical debugging tool for ML failures, we introduce \tool, an open-source, 
testing tool for ML-based software systems. Rather than relying on isolated checks or manually inspecting individual artifacts, \tool ~adopts an agentic AI framework that performs cross-artifact analysis over dataset statistics, data integrity reports, train--test validation results, and model checkpoints. Specifically, given the ML workflow, including the training dataset and the model, first of all, Deepcheck is used to find out possible bugs with respect to a set of predefined properties. Then the bug report is forwarded to a stateless analysis server. Therein, specialized LLM-backed agents analyze these bug reports in parallel and a subsequent reasoning agent synthesizes their findings into a unified diagnosis, ranking the found failures in the ML pipeline by severity and confidence. Most importantly, \tool supports diverse ML modalities such as tabular, vision, and NLP tasks, and moves beyond traditional testing tools by not only detecting potential failures, but also explaining why they occur and what practitioners can do to fix them.

\section{\tool testing framework}
\label{sec:framework}



\begin{figure*}[t]
\centering
\includegraphics[width=1.0\textwidth]{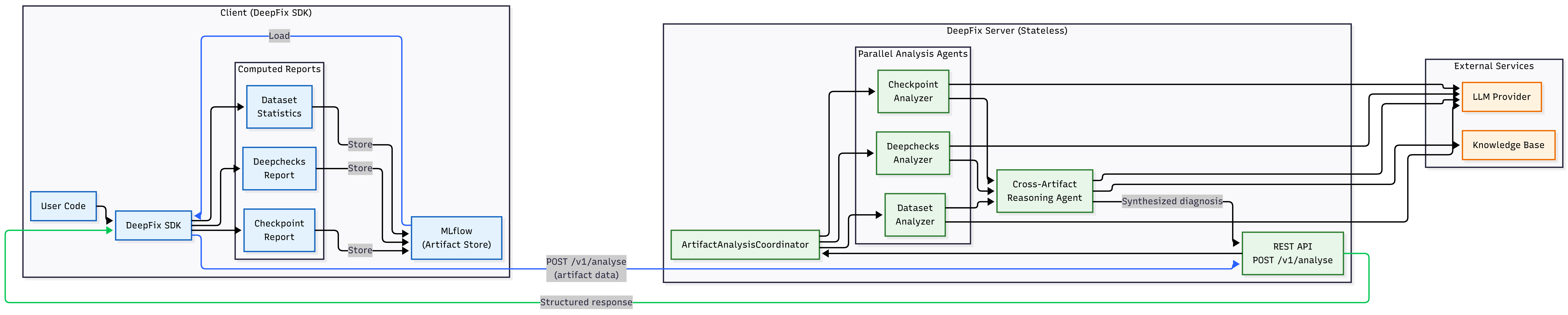}
\caption{High-level architecture of \tool. The SDK (client) performs testing, stores and filters the unnecessary details, and sends them to the server via a POST request. The stateless server dispatches them to three parallel analysis agents, whose findings are synthesized by a cross-artifact reasoning agent backed by an LLM provider and a knowledge base of best practices.}
\label{fig:arch}
\end{figure*}

In this work, we focus on testing the entire workflow of the ML system. This includes testing the \textit{training data} and the ML algorithm (i.e., the implementation of the algorithm) that is trained on the data. Together, this is termed as an ML workflow. In the rest of the paper, we will follow this terminology.
We start by giving a brief description of the architecture of the \tool. We develop it as a \textit{client-server} architecture wherein we have two separate stages. On the client side, the testing of the given ML workflow is done by Deepchecks with respect to some specific properties. Typically, these are \textit{functional properties} that correspond to the statistics of the data and the functionality of the algorithm. The bug report generated by the Deepchecks cannot be easily interpreted by the non-domain experts. Therefore, to make it more meaningful and to find out the possible fix of these bugs, this report is then sent to the server side, where an agentic-AI-based framework is further used to generate a detailed bug report with possible fixes. Figure~(Fig.~\ref{fig:arch}) gives the arcitecture of our \tool. Next, we describe these in detail below.

\subsection{The testing approach}

\tool frames ML debugging as a \emph{cross-artifact diagnostic testing} problem.
Rather than monitoring a single metric or running isolated data checks, the system inspects multiple checks simultaneously and reasons about their interactions. The testing process unfolds in two phases.

\subsubsection{Phase~1: ML Workflow Testing (client-side)}

\begin{table*}[t]
\centering
\caption{Overview of Diagnostic Checks for Tabular Data Analysis from Deepchecks\cite{chorev2022deepcheckslibrarytestingvalidating}}
\label{tab:tabular-checks}
\begin{small}
\begin{tabular}{p{0.3\linewidth} p{0.3\linewidth} p{0.3\linewidth}}
\toprule
\textbf{Data Integrity} & \textbf{Train Test Validation} & \textbf{Model Evaluation} \\
\midrule
Mixed Nulls & String Mismatch Comparison & Model Info \\
Percent Of Nulls & Train Test Samples Mix & Segment Performance \\
String Mismatch & New Label & Unused Features \\
String Length Out Of Bounds & Datasets Size Comparison & Performance Bias \\
Feature Label Correlation & New Category & Simple Model Comparison \\
Feature Feature Correlation & Index Leakage & Weak Segments Performance \\
Is Single Value & Feature Label Correlation Change & Model Inference Time \\
Columns Info & Multivariate Drift & Regression Systematic Error \\
Class Imbalance & Label Drift & Single Dataset Performance \\
Outlier Sample Detection & Date Train Test Leakage Overlap & Train Test Performance \\
Identifier Label Correlation & Date Train Test Leakage Duplicates & ROC Report \\
Conflicting Labels & Feature Drift & Regression Error Distribution \\
Data Duplicates & & Confusion Matrix Report \\
Special Characters & & Boosting Overfit \\
Mixed Data Types & & Prediction Drift \\
& & Calibration Score \\
& & Multi Model Performance Report \\
\bottomrule
\end{tabular}
\end{small}
\vspace{-1em}
\end{table*}

In this stage, the \tool checks the training data and the training algorithm.
Firstly, several function properties pertaining to dataset statistics, data integrity checks, and model checkpoints are checked and 
logged to an MLflow tracking server \cite{mlflow_developments}. Specifically, the \tool provides typed dataset wrappers for various machine learning tasks across supported modalities, including image classification, object detection, semantic segmentation, tabular data analysis, and natural language processing.
They accept standard data structures (PyTorch DataLoaders or pandas DataFrames). When the user calls \texttt{client.ingest(...)},
\tool populates a dense diagnostic feature set by executing a sequence of computations and an exhaustive suite of up to 80 distinct checks (tailored by modality: 44 for tabular as illustrated in Table~\ref{tab:tabular-checks}, 19 for NLP, and 17 for computer vision). This can be categorized as follows.

\begin{enumerate}
\item \textbf{Dataset statistics.} Given the ML workflow, the \tool first of all extracts the necessary metadata such as sample counts, class distributions, feature types, and basic distribution summaries from the training dataset.
\item \textbf{Data integrity checks (27 possible checks).} The SDK of \tool then runs a configurable suite identifying structural issues, including mixed nulls detection, string mismatches, feature-to-feature/identifier-to-label correlations, outlier sample detection, and conflicting labels.
\item \textbf{Train/test validation (22 possible checks).} Furthermore, Deepchecks is applied to evaluate the split quality by detecting train-test sample mixing, index/date leakage, dataset size discrepancies, and assessing multivariate, feature, and label drift.
\item \textbf{Model performance checks (31 possible checks).} After the dataset is checked, in the next step, if the training algorithm is provided, then \tool also evaluates inference performance, localizes the weak segments, checks for performance bias, compares single-dataset vs. split performance, and finally checks for overfit and calibration errors.
\item \textbf{Model checkpoint metadata.} The aforementioned steps check the components of the ML workflow. To get the meaningful information from the results of all the checks, \tool extracts architecture summaries, parameter counts, and the model's Python object docstring. This further enables us to combine all the relevant failures in the later stages to suggest possible fixes.
\end{enumerate}

This design keeps all heavy computation on the client side. Furthermore, given the bug reports generated by Deepchecks, we use \tool to filter out unnecessary details and generate a more structured report that the server can analyze without requiring direct access to raw datasets or training infrastructure. We call this structured report as \textit{artifact data}.

\subsubsection{Phase~2: Multi-agent analysis (server-side)}

Upon receiving an analysis request containing the client's artifact data, the server dispatches three specialized LLM-backed analysis agents that run in parallel to generate a more meaningful report. Specifically, in this phase, our tool analyzes the given bug report to produce a more interpretable and detailed report containing possible fixes along with their severity. The three agents perform the following tasks.

\begin{enumerate}
\item \textbf{Dataset Analyzer}: examines dataset statistics to surface data-quality issues such as class imbalance, anomalous feature distributions, or missing labels.
\item \textbf{Deepchecks Analyzer}: interprets the Deepchecks reports, highlighting integrity violations and distribution shifts between training and validation sets.
\item \textbf{Model Checkpoint Analyzer}: validates checkpoint integrity, checks configuration consistency, and assesses deployment readiness.
\end{enumerate}

\begin{figure*}[htbp]
\centering
\includegraphics[width=\textwidth]{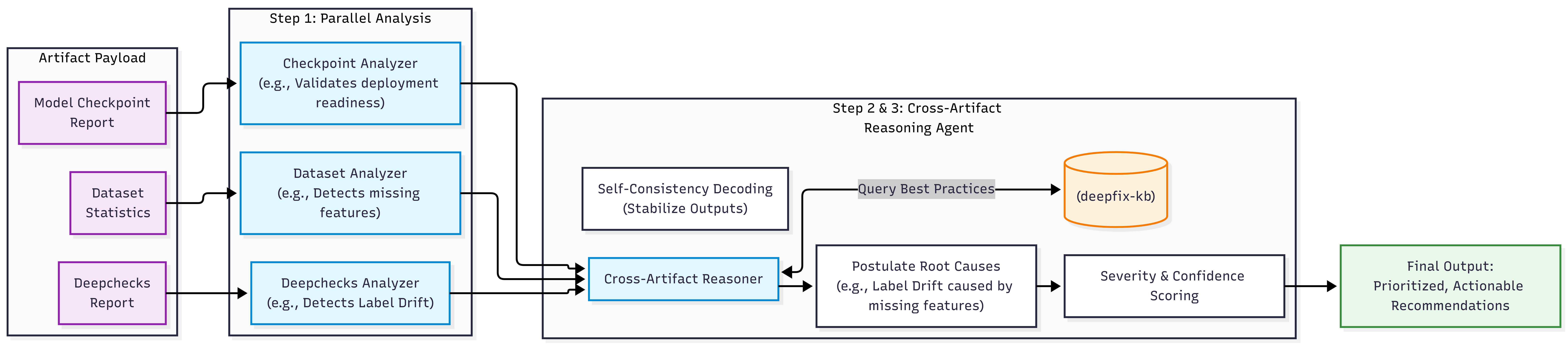}
\caption{The \tool Cross-Artifact Reasoning Procedure (Phase 2). Artifacts generated by parallel analyzers are aggregated, hypothesis formulated using the knowledge-base, and finalized through self-consistency decoding to output actionable recommendations.}
\label{fig:reasoning}
\end{figure*}

After the parallel analysis completes, a sequential cross-artifact reasoner synthesizes the findings. This synthesis follows a structured inference procedure as illustrated in Fig.~\ref{fig:reasoning}). This is done in the following three steps.
\begin{enumerate}
\item \textbf{Aggregation of Findings:} In this step, the agent consolidates the localized findings from the respective analyzers, correlating the bugs that are being found. For instance, a high misclassification rate recognized by the \textit{Deepchecks Analyzer} with root causes being a severe class imbalance can be highligthed by the \textit{Dataset Analyzer}.
\item \textbf{Hypothesis Generation \& Grounding:} After aggregating the correlated findings against canonical failure modes, an agent postulates potential root causes (referred as hypothesis). During this phase, it queries the \texttt{deepfix-kb} retrieval layer to ground these findings in best practices, ensuring the diagnosis is supported by established ML literature. Debugging best practices are fetched from the internet using \textit{Sonar} from Perplexity AI~\cite{perplexity}. \textit{Sonar} is a leading web-search grounded chat completions model.
\item \textbf{Ranking \& Uncertainty Estimation:} In this stage, each of the findings is scored based on the severity, i.e.,  considering the potential impact on model deployment and confidence. 
\end{enumerate}

Finally, this phase yields a unified, natural-language diagnosis with a prioritized, ranked list of concrete next steps. By transforming fragmented artifacts into a cohesive narrative, this two-phase approach makes \tool an \emph{active auditor}. Instead of merely displaying metrics, it shows \emph{why a model underperforms} and most importantly,  \emph{what can be done about it}, extending beyond the capabilities of existing MLOps platforms like MLflow~\cite{mlflow_developments} or wandb~\cite{wandb}.

\subsection{Extension Points}

\tool is designed to be extended by both its maintainers and the community:

\begin{itemize}
\item \textbf{New Analysis Agents.} Developers can add a new agent in four steps: (1)~implement an \texttt{Agent} subclass, (2)~register it in the \texttt{ArtifactAnalysisCoordinator}, (3)~wire any new artifact types or knowledge sources, and (4)~add tests and documentation.
\item \textbf{New Properties.} The Pydantic-based data model makes it straightforward to define new properties schemas, for instance, for reinforcement-learning rollouts or time-series forecasting logs, and have them flow through the existing pipeline.
\item \textbf{Custom LLM providers.} The provider-agnostic abstraction layer allows plugging in fine-tuned or domain-specific language models, including models trained on the organization's own debugging data.
\item \textbf{Knowledge base expansion.} The \texttt{deepfix-kb} package can be extended with curated best-practice documents covering data quality, training strategies, and architecture guidelines.
\item \textbf{Mitigating hallucination risk.} Employ self-consistency decoding to generate multiple reasoning paths and aggregate them to produce a stable consensus diagnosis.
\end{itemize}

\section{Validation}
\label{sec:validation}

\begin{table*}[t]
\centering
\caption{Industrial Food Waste Quantification: \tool Results \& Analysis}
\label{tab:cv-checks}
\begin{small}
\begin{tabular}{p{0.5\linewidth} p{0.5\linewidth}}
\toprule
\textbf{Finding} & \textbf{Action} \\
\midrule
\textbf{Critical data partitioning failure causing unrepresentative test set} \par \textit{Evidence. Combined evidence from Deepchecks: Label drift check failed with Cramer's V score of 0.92 (far exceeding 0.15 threshold) and 75\% of test set labels were not present in training} & \textbf{Immediately halt model development and recreate the train-test split using proper stratified sampling techniques}. \par \textcolor{gray}{The test set is fundamentally invalid for evaluation due to severe Label distribution mismatch and Leakage, making any model performance metrics meaningless} \\
\textbf{Systematic differences in image acquisition conditions between datasets} \par \textit{Evidence. Multiple image property drift, failures: Brightness (KS=0.42), RMS Contrast (KS=0.5), Red Intensity (KS=0.83), Green Intensity (KS=0.82), Blue Intensity (KS=0.96)} & \textbf{Standardize image collection protocols and apply normalization techniques to align visual characteristics} \par \textcolor{gray}{Large differences in brightness, contrast, and color properties will cause models to learn dataset-specific artifacts rather than generalizable features} \\
\bottomrule
\end{tabular}
\end{small}
\vspace{-1em}
\end{table*}

Although a large-scale empirical study is currently in planning, \tool has already undergone initial validation through stakeholder interviews, live demonstrations, and in an industrial project. We have tested \tool in a consulting project as part of Delcaux Labs activities where it successfully identified train-test partitioning mistakes in the context of image-based food waste quantification, see Table~\ref{tab:cv-checks}. \tool correctly identified that the image collection protocol was not standardized. Viewing hundreds of images to come to this finding would have required extensive effort. We are unable to disclose the name of the involved company due to confidentiality.
Apart from the already performed validation, we further plan to extend.

For instance, to robustly isolate \tool's value, the study will compare its diagnostic performance against established baselines. Additionally, we will conduct ablation studies to quantify the impact of the knowledge base, compare cross-artifact versus single-artifact synthesis, and evaluate the LLM-based reasoning against rigid rule-based heuristics. Secondary evaluation metrics will include an \emph{actionability score} (expert review of proposed remedies), \emph{hypothesis ranking quality} (Mean Reciprocal Rank of the correct root cause), and \emph{computational efficiency} (offline profiling cost and online reasoning latency). Furthermore, to ensure the tool stands up to regular MLOps workloads, \tool will undergo scale testing. Specifically, we plan to track:
\begin{itemize}
    \item \textbf{Efficiency metrics:} Maintaining a mean-time-to-diagnosis (MTTR) under 60 seconds for end-to-end flows, with a target cache hit rate over 40\%.
    \item \textbf{User impact:} Conducting a 30-participant study to measure usability observing the real-world adoption rate of the generated recommendations.
\end{itemize}

\subsection{Limitations}
\label{sec:limitations}
While \tool accelerates ML debugging, it has limitations common to LLM-backed systems.
First, the multi-agent pipeline introduces higher latency and compute costs than traditional, heuristic-based linters. Future work will explore caching to mitigate this.
Second, deploying an active auditor requires sending artifact data (e.g., distribution statistics or checkpoints) to LLM providers, raising potential data privacy and compliance concerns for proprietary models. To alleviate this, \tool supports on-premise execution by connecting to self-hosted language models, ensuring sensitive artifacts never leave the organizational perimeter.
Third, despite our grounding strategy through \texttt{deepfix-kb}, LLM hallucinations and prompt sensitivity remain risks. While we employ constrained evidence aggregation to stabilize outputs, human-in-the-loop validation continues to be a necessity.
Finally, \tool's effectiveness depends on the quality of its knowledge base; continuous effort is needed to curate and update these best practices which are currently fetched from Sonar (i.e. Perplexity AI).

\section{Conclusion}
\label{sec:conclusion}

We introduced \tool, an open-source diagnostic assistant that addresses the gap between complex ML failures and practical debugging tools.
By combining already existing state-of-the-art testing tool with a multi-agent analytical pipeline, \tool continuously audits artifacts to hypothesize the root causes of model underperformance, yielding actionable, knowledge-backed recommendations. Based on the initial validations with the industry projects, we already found that \tool can be quite impactful in the debugging tasks of ML workflow. The modular structure of our tool \tool further enables easy extensions of integrating new properties, testing approaches, amongst others. Therefore, in the future, we plan to further extend, considering several aspects including rigorously benchmarking \tool's scientific accuracy, including ablations against siloed tools, and confirming its operational efficiency across diverse datasets and modalities.

\section*{Tool Availability}
\tool is available under the Apache 2.0 license at the project repository: \url{https://github.com/delcaux-labs/deepfix/}. All of its dependencies are open-source.

\section*{Acknowledgments}
Funded by the Federal Ministry of Research, Technology and Space under the funding code “KI-Servicezentrum Berlin-Brandenburg” 16IS22092.
Responsibility for the content of this publication remains with the author.

\bibliographystyle{IEEEtran}
\bibliography{ref}

\end{document}